\documentclass[aps,pre]{revtex4}
\usepackage{epsfig}
\def\D{\displaystyle}

\begin{document}
\title{Light scattering from cold rolled aluminum  surfaces}
\author{Damien Vandembroucq$^\dag$, Annie Tarrats$^\ddag$, Jean-Jacques Greffet$^\ddag$,
St\'ephane Roux$^\dag$ and Franck Plourabou\'e$^*$}
\affiliation{
$^\dag$ Unit\'e mixte  CNRS/Saint-Gobain ``Surface du Verre et Interfaces'',
    93303 Aubervilliers Cedex, France\\
$^\ddag$ Laboratoire d'Energ\'etique Mol\'eculaire et Macroscopique, Combustion\\ 
Ecole Centrale Paris, 92295 Chatenay Malabry Cedex, France \\
$^*$ Institut de M\'ecanique des Fluides,
All\'ee du Professeur Camille Soula ,
31400 Toulouse, France}


\begin{abstract} 
We present experimental light scattering measurements from aluminum
  surfaces obtained by cold rolling. We show that our results are
  consistent with a scale invariant description of the roughness of
  these surfaces. The roughness parameters that we obtain from the
  light scattering experiment are consistent with those obtained from
    Atomic Force Microscopy measurements.
\end{abstract}
\maketitle

\begin{figure}
\begin{center}
\epsfig{file=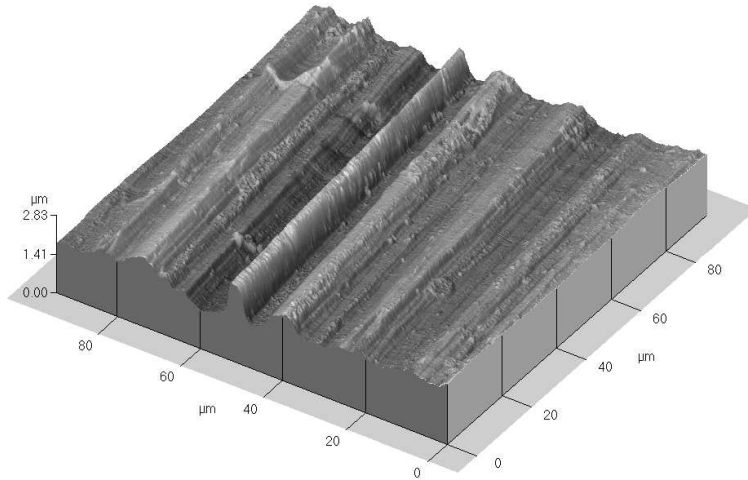,angle=-90,width=11cm}
\end{center}
\caption{AFM image of 512$\times$512 points of the aluminum alloy
 sheet surface. This image has been obtained by Plourabou\'e and Boehm\cite{laminage}
 in contact mode on a Park Scientific AFM using a long range scanner
 (100 $\mu$m lateral travel and 5 $\mu$m vertical travel). 
The height standard deviation has been measured to be $\sigma= 2.5\;\mu{\rm m}$.}
\label{profilometry}
\end{figure}

Since an early paper by Berry in 1979\cite{Berry}, the study of wave
scattering from self-affine (fractal) surfaces has become very active,
see
Ref.\cite{Jaggard,Shepard,MacSharry,Lin,Chen,Sheppard,Sanchez,Sanchez2,Zhao}
for recent references.  Most of these papers consist in numerical
simulations; apart from the early works of Jakeman {\it et al}
\cite{Jakeman86,Jakeman88} very few theoretical results have been
published; the same statement stands for experimental results while
lots of real surfaces\cite{EB,Meakin,laminage} have been shown to obey
scale invariance.  Here we try and test experimentally recent
theoretical expressions obtained for the scattering of a scalar wave
from a perfectly conducting self-affine surface \cite{SRVPRE}.  We
report scattering measurements of an s-polarized electromagnetic wave
(632.8 nanometers) from a rough aluminum alloy plate (Al 5182).  The
latter was obtained by industrial cold rolling.  As presented in
Fig. \ref{profilometry} taken from Ref.  \cite{laminage} by
Plourabou\'e and Boehm, the rolling process results in a very
anisotropic surface, the roughness being much smaller along the
rolling direction than in the orthogonal one.  From Atomic Force
Microscopy (AFM) measurements with a long range scanner the authors
could establish the scale invariant character of the roughness: the
surface was found to be self-affine between a few tens of nanometers
and about fifty micrometers. At the macroscopic scale, they measured the height standard deviation (RMS roughness) to be $\sigma=2.5\;\mu{\rm m}$.

Let us briefly recall that a profile or a one-dimensional surface is
said to be self-affine if it remains statistically invariant under the
following transformations:
$$
x \to \lambda x \;, \quad
z \to \lambda^{\zeta} z \;.
$$
where the parameter $\zeta$ is the roughness exponent. A direct
consequence of this scale invariance is that when measured over a
length $d$ geometrical quantities such as a roughness $\sigma$ or a
slope $s$ are dependent on this length $d$:
$$
\sigma(d) \propto d^{\zeta}\;, \quad s(d) \propto  d^{\zeta-1}
$$

The roughness exponent which characterizes the autocorrelation
function is however not sufficient to give a complete characterization
of the statistics of the surface roughness. The latter also requires
an amplitude parameter. In the context of light scattering, one can
for example normalize the geometrical quantities with their value over
one wavelength:
$$
\sigma(d) = \sigma(\lambda) \left(\frac{d}{\lambda}\right)^{\zeta}\;, 
\quad s(d)= s(\lambda) \left(\frac{d}{\lambda}\right)^{\zeta-1}
$$
We will see in the following that the value of the slope $s(\lambda)$
is the crucial numerical parameter when dealing with scattering from
self-affine rough surfaces. Note finally that the scale invariance of
real surfaces roughness can only extend over a finite domain. The
upper cut-off allows to define a macroscopic roughness, the lower one
allows to define a local slope in every point.  This scaling invariant
formalism has been shown to be relevant to describe varied surfaces
such as the ones obtained by fracture\cite{EB}, growth or deposition
processes \cite{Meakin}.

\begin{figure}
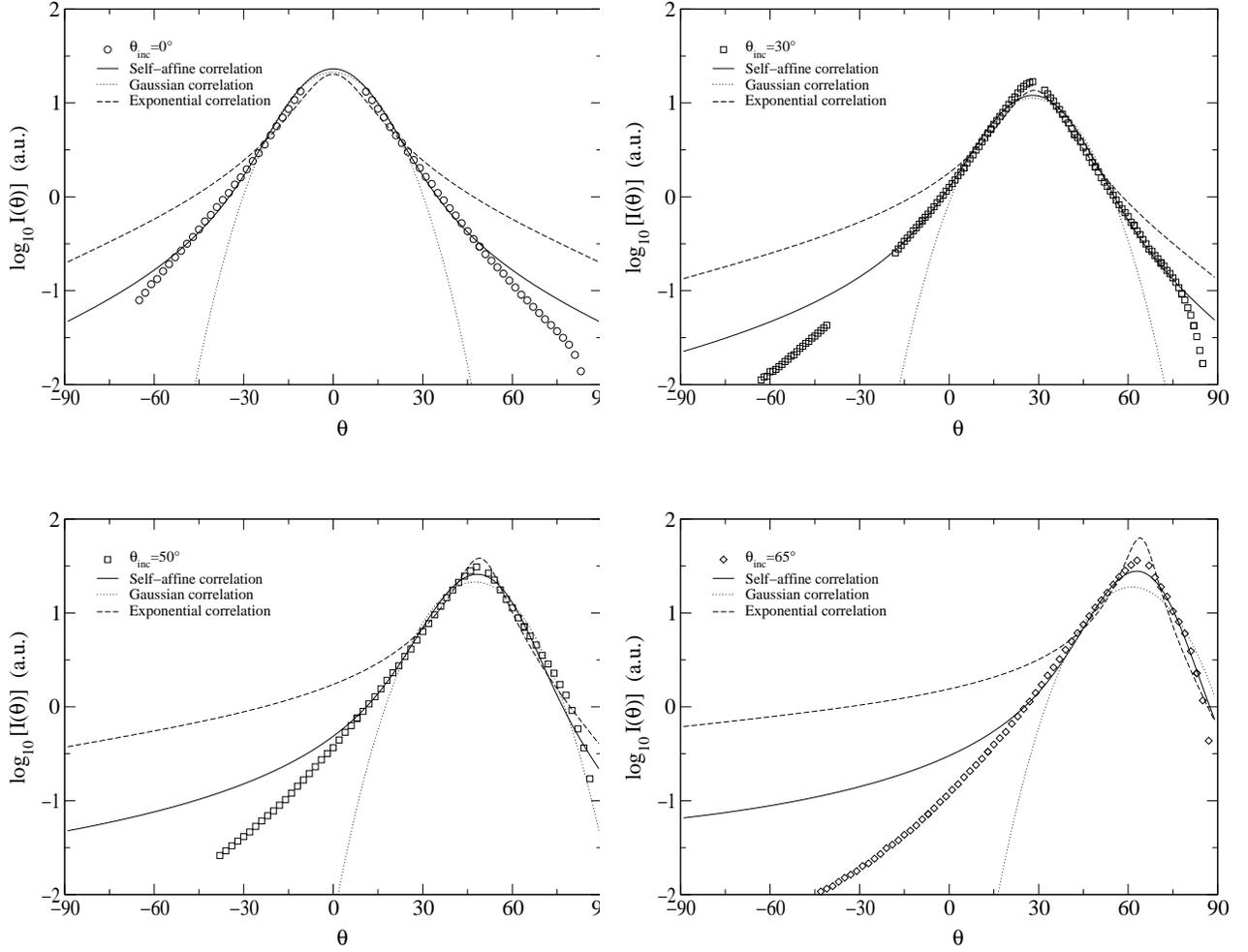

\begin{tabular}{cc}
\vspace{0.75cm}
\epsfig{file=fig_inc0.eps,width=6.truecm,angle=-90}
&\epsfig{file=fig_inc30.eps,width=6.truecm,angle=-90}\\
\vspace{0.75cm}
\epsfig{file=fig_inc50.eps,width=6.truecm,angle=-90}
&\epsfig{file=fig_inc65.eps,width=6.truecm,angle=-90}\\
\end{tabular}
\caption{Scattered intensity measurements obtained at incidence angles
$\theta_0=0$, $30$, $50$ and $65$ degrees respectively. The
experimental results are shown in symbols. The solid/dotted/dashed
lines correspond to the expressions obtained for a Kirchhoff
approximation in case of self-affine/Gaussian/exponential correlations
respectively.}
\label{angle0}
\end{figure}

We performed our measurements on a fully automated scatterometer (see
ref. {\cite{Greffet-PhD,Greffet-manip}} for a full description).  The
set-up is designed for the measurement of the bidirectional scattering
distribution function.  The source is a Helium-Neon laser of
wavelength $\lambda=632.8$ nm, the beam passes through a mechanical
chopper and is submitted to a spatial filtering before reaching the
sample.  The latter is placed on a rotating plate which allows to vary
the incident angle.  The scattered light is collected by a converging
lens and focussed on a photomultiplier.  This detection set-up is
placed on an automated rotating arm. Note that the shadow of the
photomultiplier imposes a blind region of $\pm 11$ degrees around the
back-scattering angle.  Two polarizers allow us to select the
polarization directions of both incident and scattered lights. The
output signal is filtered by a  lock-in
amplifier and processed by a micro-computer. We used a frequency
$f=700 {\mathrm Hz}$ and a time constant $\tau= 1 {\mathrm s}$.  The
surface being highly anisotropic, the result is {\it a priori} very
sensitive to the orientation of the surface. 
In order to select
properly one of the two main directions of the surface, we placed a
vertical slit in front of the photomultiplier. 
 This allows to reduce the effects of possible misorientation of the
sample.
The results of the scattering measurements obtained for incidence
angles 0, 30, 50 and 65 degrees are displayed in semi-log scale on figure
\ref{angle0}.

How does the scale invariance of the roughness affect the angular
distribution of the scattered light?  The comparison of experimental
light scattering data with theoretical models still remains a delicate
matter. A key point is obviously to give a proper
description of the statistical properties of the surface
roughness. When testing new models or approximations, it is usual to
design surfaces of controlled Gaussian autocorrelation function (this
is for example possible by illuminating photosensitive materials with
a series of laser speckles \cite{Knotts93,Knotts94,Calvo}).  In the following we want to
test the consistency of our scattering measurements with the roughness
analysis. We perform this test {\it via} a very crude approximation:
we consider the surface to be one-dimensional and perfectly conducting.
We then compare our experimental results with analytical predictions
obtained in the context of a simple Kirchhoff approximation
corresponding to Gaussian, exponential and self-affine correlations.

Although lots of studies have been published about scattering from
scale invariant surfaces in the last twenty years, very few analytical
results can be found in the literature. The main results are due to
Jakeman and his collaborators \cite{Jakeman86,Jakeman88} who showed
that the angular distribution of the intensity of a wave scattered
from a self-affine random phase screen could be written as a L\'evy
distribution. In a similar spirit, some of us studied very
recently\cite{SRVPRE} the case of scattering from self-affine surfaces
and found in the context of a Kirchhoff approximation:

\begin{mathletters}
    \label{sa}
\begin{eqnarray}
    \label{MDRC-final}
    \left< \frac{\partial R}{\partial \theta} \right>
    &=&
    \frac{s(\lambda)^{-\frac{1}{\zeta}}a^{-(\frac{1}{\zeta}-1)}}{\sqrt{2}\, \cos\theta_0 }
     \frac{
    \cos\frac{\theta+\theta_0}{2}}{\cos^3\frac{\theta-\theta_0}{2}}  \\
\nonumber
 &&\times {\cal L}_{2\zeta}\left(\frac{ \sqrt{2} \tan\frac{\theta-\theta_0}{2} }
{ 
a^{\frac{1}{\zeta}-1} 
s(\lambda)^{\frac{1}{\zeta}}}
  \right),  
\end{eqnarray}
where 
$a = 
2 \pi \sqrt{2}
\cos\frac{\theta+\theta_0}{2}\cos\frac{\theta-\theta_0}{2}$,
and ${\cal L}_{\alpha}(x)$ is the centered symmetrical L\'evy stable
distribution of exponent $\alpha$ defined as
\begin{eqnarray}
    \label{Levy-distribution-1}
    {\cal L}_\alpha(x) &=& \frac{1}{2\pi} \int^\infty_{-\infty} dk \; 
                 e^{ikx} e^{-\left| k \right|^\alpha}. 
\end{eqnarray}
\end{mathletters}

Note that the form of this analytical result does not depend on the
value of the global RMS roughness $\sigma$ in contrast to the case of
a Gaussian correlated surface. 
The scattering pattern is centered around the specular direction with an
angular width $w$ which scales as

$$
w \simeq s(\lambda)^{\frac{1}{\zeta}}
$$
It is worth mentioning here that in the context of this simple
Kirchhoff approximation, the crucial geometrical parameter to consider
is the slope over the scale of one wavelength $s(\lambda)$: the
angular distribution of the scattered intensity is mainly controlled
by this ``local'' parameter and does not depend on the value of the
global RMS roughness. The latter will only come back into the game if
one goes beyond a single scattering approximation.

Using the complete set of experimental scattering data, we performed a
numerical fitting procedure for the expression (\ref{MDRC-final}) and
for the expressions obtained with Gaussian or exponential
correlations.  The latter have been derived in the case of very rough
surfaces (see Appendix for details of the expressions and the
derivation). The fitting procedure consisted in a numerical
minimization of the quadratic distance between the data and the tested
expression in logarithmical scale. The free parameters are an
amplitude parameter (which is simply an additive constant in
logarithmic scale) and two geometrical parameters: the roughness
exponent $\zeta$ and typical slope over the wavelength
$s(\lambda)$. In the case of gaussian or exponential correlation there
is only one geometrical parameter which is an equivalent slope
$\sigma/\tau$ or $2\pi\sigma^2/\lambda\tau$ respectively. Note that
the same parameters are used for the whole set of experimental data
gathering four different incidence angles.


\vspace{0.2cm}

In order to get rid of shadowing and multiple scattering effects, we
restricted the fitting procedure to a region of $\pm$ 50 degrees
around the incidence angle. In this region we can see on figure
\ref{angle0} that there is a good
agreement with the expression (\ref{sa}) which has been obtained with
a roughness exponent $\zeta=0.78$ and a typical slope over the
wavelength $s(\lambda)=0.11$. For large scattering angles  the analytical
expression systematically overestimates the scattered intensity.
We attribute this behavior to the shadowing efffects.
None of the Gaussian and exponential correlations can give a
comparable result. In the Gaussian case, we obtain $\sigma/\tau=0.08$ 
and  in the exponential case  $2\pi\sigma^2/\lambda\tau=0.10$. 

Beyond this direct comparison of the different prediction for the
angular distribution of the scattered intensity, we try also to
compare the geometrical parameters that we obtained with direct
roughness measurements performed by Atomic Force Microscopy. We imaged
an area of size 2.048 $\mu$m $\times$ 2.048 $\mu$m with a lateral step
of 4 nm.  From these roughness measurements we compute the typical
height difference $\Delta z$ between two points as a function of the
distance $\Delta x$ separating the two points. This quantity is
obtained {\it via} a quadratic mean over all possible couples of
points separated by a given distance $\Delta x$.  In case of
self-affine, Gaussian or exponential correlations, we expect
respectively:

\begin{eqnarray}
\vspace{2pt}
\Delta z_{\mathrm sa}&= &\lambda s(\lambda)
 \left(\frac{\Delta x}{\lambda} \right)^\zeta \;,\\
\vspace{2pt}
\Delta z_{\mathrm Gauss}&= &\sigma \sqrt{2} 
\sqrt{1-\exp(-\frac{\Delta x^2}{\tau^2})} \;,\\
\Delta z_{\mathrm exp}&= &\sigma \sqrt{2} 
\sqrt{1-\exp(-\frac{\Delta x}{\tau})}\;.
\end{eqnarray}

\begin{figure}[t]
\begin{center}
\epsfig{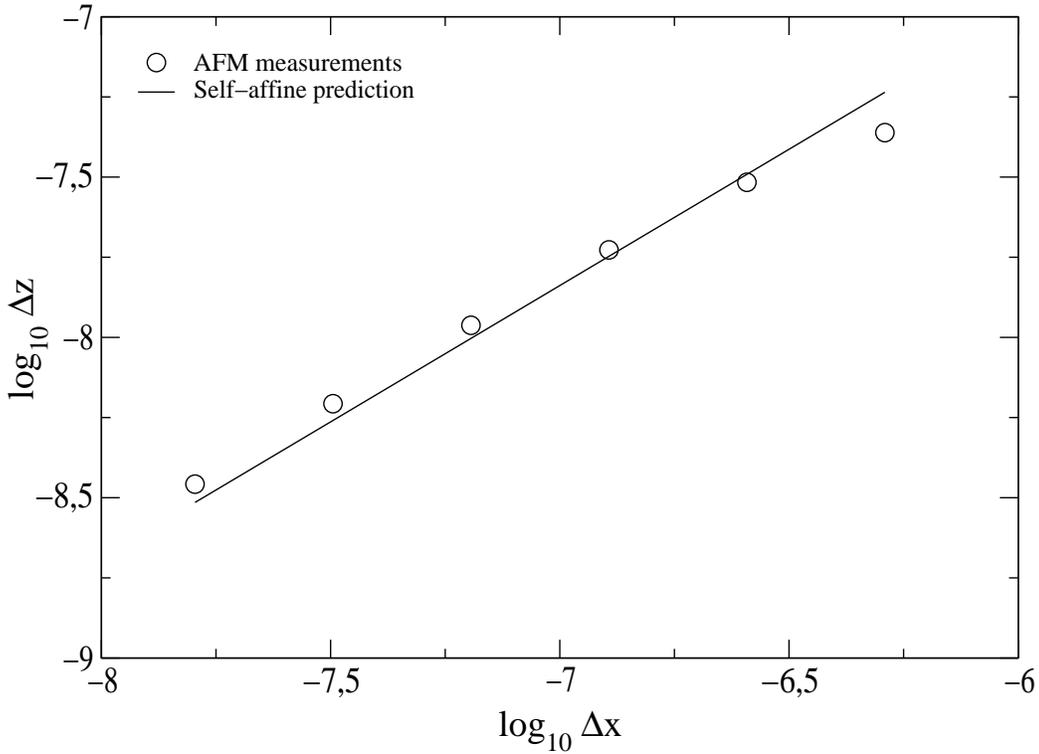}
\end{center}
\caption{Roughness analysis computed from AFM measurements (circles)
compared with predictions obtained {\it via} a fit of the angular
scattered intensity distribution assuming self-affine
correlations. The slope of the line is $\zeta=0.78$}
\label{comparaison}
\end{figure}

We show in Fig.\ref{comparaison}  the results of the  roughness
analysis and the predictions corresponding to the
self-affine correlations.  Both the value $\zeta=0.78$
of the roughness exponent and the slope over one wavelength
$s(\lambda)=0.11$ that we obtain from the scattering measurements seem
to be consistent with the experimental roughness data.  Note that the
hypothesis of exponential and Gaussian correlations would have lead to 
 power laws of exponents 0.5 and 1 respectively since we consider
horizontal distances $\Delta x$ about the wavelength which are far
smaller than the expected correlation lengths.

These first results can be considered as very promising: let us recall
that we assumed the surface to be purely one-dimensional and perfectly
conducting and that we used a basic Kirchhoff approximation,
neglecting all shadowing or multiple scattering effects...  Refining
the modeling of shadowing or multiple scattering in the specific case
of self-affine surfaces could allow to design a valuable tool to
measure the geometrical parameters describing self-affine
surfaces. This experimental study also makes clear that self-affine
correlations can be a relevant formalism to describe the optical
properties of real surfaces. Beyong classical optical phenomena this
could be also of great interest in the context of the recent studies
\cite{Carminati,Greffet-Carminati} modeling thermal emission properties of rough
surfaces.

\vspace{0.25cm}

\appendix

\section*{}
We derive in this appendix the expression of the scattering
cross-section in the framework of the Kirchhoff approximation for a
one-dimensional very rough surface.

In the following we consider the scattering of $s$-polarized
electromagnetic waves from a one-dimensional, rough surface
$z=\zeta(x)$. The height distribution is supposed to be gaussian of standard deviation $\sigma$ and
the two-points statistics is characterized by the autocorrelation
function $C(v)$. The pulsation of the wave is $\omega$, the wave number is $k$, the incidence
angle is $\theta_0$, the scattering angle is $\theta$.

Following Maradudin {\it et al} \cite{AnnPhys} the Kirchhoff approximation gives for the scattering cross-section $\partial R/\partial \theta$ from a rough surface of infinite lateral extent:

\begin{mathletters}
  \label{MDRC-mod}
  \begin{eqnarray}
    \left< \frac{\partial R_s}{\partial \theta} \right>
    &=&
    \D{   \frac{\omega}{2\pi c} \frac{1}{\cos\theta_0}
      \left( \frac{\cos \left[(\theta+\theta_0)/2\right]}
        {\cos \left[(\theta-\theta_0)/2\right]} \right)^2} {\cal I}(\theta,\theta_0)
     \;,
  \end{eqnarray}
  where 
  \begin{eqnarray}
    \label{Omega} 
    {\cal I}(\theta,\theta_0)&= &\int^{\infty}_{-\infty}dv\; 
\exp\left\{ik(\sin\theta-\sin\theta_0)v  \right\} \Omega(v) \;, \\
    \Omega(v) &=&\D{ \left< \exp\left\{-ik[\cos\theta+\cos\theta_0]\,
        \Delta\zeta(v)\right\}\right>\;.}
  \end{eqnarray}
\end{mathletters}
 Note that the statistical properties of the profile function,
$\zeta(x)$, enters Eqs.~(\ref{MDRC-mod}) only through
$\Omega(v)$. With the knowledge of the autocorrelation function
$C(v)$ the distribution of the height differences
$\Delta\zeta(v)=\zeta(x+v)-\zeta(x)$ can be written:

\begin{eqnarray}
\label{distridiff}
P\left(\Delta\zeta,v\right)=
\frac{1}{2\sigma \sqrt{\pi} \sqrt{1-C(v)}}
\exp\left[
\frac{-\Delta\zeta^2}{4\sigma^2 \left[1-C(v)\right]}
\right]\;.
\end{eqnarray}

This leads immediately to:
\begin{equation}
 \Omega(v) =\D{  \exp\left\{ -k^2 \sigma^2(\cos\theta+\cos\theta_0)^2 
 \left[ 1-C(v) \right] \right\}\;.}
\end{equation}
In case of a very rough surface, we have $k^2 \sigma^2 \ll 1$ (in our
experimental case, $\sigma =2.5 \mu{\rm m}$ and $\lambda =632.8$ nm so
that $k^2 \sigma^2 \simeq 600$) and and the only $v$ to really
contribute to the integral are in the close vicinity of zero. We can
then replace $C(v)$ by the first terms of its expansion around zero.
Consider the gaussian and exponential cases
\begin{equation}
C_{G}(v)=\exp\left(-\frac{v^2}{\tau^2}\right) \;,\quad
C_{exp}(v)=\exp\left(-\frac{v}{\tau}\right) \;.
\end{equation}
where $\tau$ is by definition the correlation length, this leads to:

\begin{eqnarray}
\Omega_G(v)&=&
\exp\left[- k^2 \left(\cos\theta+\cos\theta_0\right)^2
\alpha^2 v^2 \right] \;,
\\
\Omega_{exp}(v)&=&
\exp\left[-k^2 \left(\cos\theta+\cos\theta_0\right)^2
\alpha \sigma |v| \right]
\;.
\end{eqnarray}


Simple algebra leads finally to

 \begin{eqnarray}
    \left< \frac{\partial R_s}{\partial \theta} \right>_G
    &=&
    \D{   \frac{k}{4 \alpha \sqrt{\pi}\cos\theta_0}
       \frac{\cos \left[(\theta+\theta_0)/2\right]}
        {\cos^3 \left[(\theta-\theta_0)/2\right]}}
\exp\left[ -\frac{1}{4\alpha^2} 
\left( \tan \frac{\theta-\theta_0}{2}
 \right)^2 \right] 
     \;, \\
    \left< \frac{\partial R_s}{\partial \theta} \right>_{exp}
    &=&
\frac{\alpha \sigma k}{\pi \cos\theta_0}
\frac{\cos^2 \left[(\theta+\theta_0)/2\right]}
{\sin^2 \left[(\theta-\theta_0)/2\right] + 4(\alpha \sigma k)^2 
\cos^2 \left[(\theta+\theta_0)/2\right] \cos^4 \left[(\theta-\theta_0)/2\right]}
  \end{eqnarray}

\end{document}